\documentstyle[multicol,aps,prl,epsf,psfig]{revtex}
\begin{document}
\title{Slow Relaxation in a Constrained Ising 
Spin Chain: a Toy Model for Granular Compaction} 
\author{Satya N. Majumdar and  David S. Dean
\\  Laboratoire de
Physique Quantique (UMR 5626 du CNRS), Universit\'e Paul Sabatier,
31062 Toulouse Cedex, France}
\maketitle

\begin{abstract}
We present detailed analytical studies on the zero temperature coarsening
dynamics in an Ising spin chain in presence of a dynamically induced field 
that favors locally the `$-$'
phase compared to the `$+$' phase. We show that the presence of such a
local kinetic bias drives the system into a late time state with
average magnetization $m=-1$. However the magnetization relaxes into this
final value extremely slowly in an inverse logarithmic fashion.
We further map this spin model exactly onto a simple lattice model
of granular compaction that includes the minimal
microscopic moves needed for compaction. This toy model
then predicts analytically an inverse logarithmic law for the growth
of density of granular particles, as seen in recent experiments
and thereby provides a new mechanism for the inverse logarithmic
relaxation. Our analysis utilizes an independent 
interval approximation
for the particle and the hole clusters and is argued to be exact at
late times (supported also by numerical simulations).    

\vskip 0.5cm


\end{abstract}

\begin{multicols}{2}
\section{Introduction}
Slow relaxation dynamics naturally occurs in systems
with {\em quenched disorder} such as spin glasses
and has remained a subject of long standing interest\cite{BCKM}.
However, systems without quenched disorder such as structural glasses
also exhibit slow dynamics. It is believed that the slow
relaxation in the latter systems is due to {\em kinetic disorder},
induced by the dynamics itself\cite{Fred}. Another important class of
systems without quenched disorder is granular material, where once
again kinetic disorders are responsible for slow relaxation.
In a recent experiment\cite{GE}, a cylinder packed loosely with glass
beads was tapped mechanically and it was found that the system
gets more and more compact with time. However, the density $\rho(t)$
compactified rather slowly with time as, $\rho (\infty)-\rho (t)\sim 1/{\log (t)}$.
How robust is this inverse logarithmic relaxation? Is it only specific
to granular systems or does this also occur in other out of equilibrium systems
in presence of kinetic disorders?

In this paper we study, in detail, the effect of kinetic disorders
on the relaxation dynamics of an Ising system quenched from
a high temperature disordered phase into the low temperature ordered phase.
In absence of kinetic disorders, the dynamics of such a system
is well understood\cite{bray}. As time progresses, domains of equilibrium
low-temperature ordered phases (consisting predominantly of up and down spins
respectively) form and grow. The average linear size of a domain
grows with time as $l(t) \sim t^{1\over 2}$ for zero temperature
nonconserved dissipative dynamics. How does disorder, {\em quenched} or {\em kinetic},
affect this simple dynamics? The effect of quenched ferromagnetic disorder on
this phase ordering kinetics has been studied extensively\cite{bray}.
Essentially the quenched disorder tends to pin the domain walls leading to
complete freezing at $T=0$. However at small nonzero temperatures, the
domains still coarsen via activated dynamics but extremely slowly as,
$l(t)\sim ({\log t})^{1/4}$\cite{bray}. The purpose of this paper is to explore
the effect of {\em kinetic disorder} on the phase ordering dynamics.
A shorter version of this paper with the main analytical results 
and their numerical confirmation has appeared
elsewhere\cite{mdg}. Here we explore the dynamics in more detail and also
elaborate its connection to other systems such as granular material and
reaction-diffusion systems. 

In this paper, we study the
zero temperature dynamics of an Ising chain in presence
of a specific type of kinetic disorder namely a dynamically induced magnetic field.
If a small uniform external field (say in the down direction)
is put on in an Ising system following a
rapid quench from infinite temperature to $T=0$, the system rapidly relaxes
into a pure state with all spins down. In this case, the symmetry between
the two ordered pure states is broken globally. Interesting physics happens
when, instead of a global external bias, the symmetry between the pure states
is broken locally by the dynamics itself. In this paper, we investigate the
effect of this particular kinetic disorder and show that this system
also gives rise to a very slow dynamics. In particular, it gives rise to
inverse logarithmic relaxation (ILR) of magnetization, very similar to
the density compaction in granular systems. Our study therefore suggests that
the ILR is a very robust phenomenon and not just limited to granular systems
or specific types of kinetic disorders.

The paper is organized as follows. In Section-II, we define our model
precisely and summarize the main results. In Section-III,
we establish the connection between our spin model and a lattice model
of granular compaction. We also show that our model can be
viewed as a new one dimensional reaction-diffusion model when
the dynamics is described in terms of the kinks between domains of opposite phases.
In Section-IV, we derive some exact results. In section-V, we analyze
the dynamics via an independent interval approximation (IIA). In Section-VI,
we argue that the IIA results become exact in the limit when
the initial volume fraction of one of the phases is small. In Section-VII,
we extend this argument to other volume fractions as well. In Section-VIII,
we present a heuristic approach to support the IIA results. Finally
we conclude in Section-IX.

\section{The Model and the overview of results}

We consider a simple
Ising spin chain with spins $S_i = \pm 1$.  Starting from a given
initial configuration the system evolves by single spin flip
continuous time dynamics.  The rate of flipping of a given spin
depends on the its neighboring spins. We denote the rate
of spin flip $S_i \to -S_i$ by $W(S_i;S_{i-1}, S_{i+1})$ where
$S_{i-1}$ and $S_{i+1}$ are the two neighboring spins.
In our model the rates are specified as follows:
\begin{eqnarray}
& & W(+;++) = W(-;--) = 0 \nonumber \\ & & W(+;-+) = W(+;+-) = W(-;+-)
= W(-;-+) = {1\over 2} \nonumber\\ & & W(+;--) = 1 \nonumber\\ & &
W(-;++) = \alpha
\label{eq:rates}
\end{eqnarray}
Note that the case $\alpha = 1$ corresponds to the usual zero temperature Glauber
dynamics\cite{glauber} which preserves the symmetry between the up and down
phases. However, if $\alpha<1$, the flipping of a down spin sandwiched between two up spins
is not as likely as the flipping of an up spin sandwiched between two down spins.
Thus $\alpha<1$ clearly breaks the symmetry between the up and down phases. However
this symmetry is broken only dynamically, i.e. not everywhere but only at
at the location of the triplets $(+,-,+)$.
Thus the
isolated `$-$' spins (surrounded on both sides by a `$+$') tend to block the
coalescence of `$+$' domains and locally favor the `$-$' spins.
One can argue that the asymptotic dynamics
at late times is similar for any $\alpha<1$. In other words, $\alpha=0$ is an attractive
fixed point. We therefore restrict ourselves only to the case $\alpha=0$.

To see the effect of the local dynamical constraint more precisely, we
derive (following Glauber's calculation for $\alpha=1$\cite{glauber})
the exact evolution equation for the magnetization $m(t)=\langle S_i
\rangle$ for the $\alpha = 0$ case:
\begin{equation}
{d\over dt}\langle S_i \rangle = - 2 P(1,-1,1),
\label{eq:em}
\end{equation}
where $P(\sigma_{i-1},\sigma_{i},\sigma_{i+1})(t) = \langle (1 + \sigma_{i-1}S_{i-1})(1 + \sigma_{i} S_{i}) (1 + \sigma_{i+1}S_{i+1})\rangle/8 $
is the three point
probability to find the sequence of spins $(\sigma_{i-1}, \sigma_i, \sigma_{i+1})$ about the site $i$
 and we have used translational invariance.  Hence
 $P(1,-1,1)(t)$ denotes the probability of the occurrence of the triplet
`$+-+$' at time $t$. We note that for the case $\alpha = 1$, $d\langle
S_i \rangle /dt = 0$ \cite{glauber}, indicating that the magnetization
does not evolve with time. In our case, due to the triplet defects
`$+-+$', the average magnetization decays with time.  If $L_{\pm}(t)$
denote the fractions of `$+$' and `$-$' spins, then using
$L_{\pm}=(1\pm m)/2$ we find from Eq. (\ref{eq:em}), $dL_{\pm}/dt =\mp
R_1(t)$ where $R_1(t)$ is the number density of the triplets of type
`$(+-+)$' per unit length,
clearly exhibiting the asymmetry generated by the `$(+-+)$'
triplets.  We also note that unlike the $\alpha = 1$ case, the
evolution equation (\ref{eq:em}) for the single point correlation
function involves two and three point correlations (via $R_1(t)$).
This hierarchy makes an exact solution difficult for  $\alpha =0$.

It is useful at this point to summarize the main results obtained in this paper.
Let us first highlight the contrast between the $\alpha=1$ (no kinetic disorder)
and the $\alpha=0$ (with kinetic disorder) cases. For $\alpha=1$, due to the
preserved symmetry between the up and down phases at all times,
the average domain size of both `$+$' and `$-$' domains grow as
$\l_{\pm}(t)\sim t^{1/2}$ at late times\cite{bray1}. Thus the average
magnetization $m(t) = (l_+ - l_-)/(l_+ + l_-)$ is a constant of
motion\cite{glauber}) and stays fixed at its initial value. In particular, if we start from
an initial state where the spins are random (infinite temperature), the initial
average magnetization is zero and stays $0$ at all subsequent times.
In contrast, for $\alpha=0$ case where the symmetry is dynamically broken,
we find the following results.

(1) For $\alpha=0$, while domains of both phases continue to grow with time,
they have different growth laws.
The average domain sizes of the $+$ phases (denoted by $l_{+}(t)$) and $-$ phases (denoted by
$l_{-}(t)$) coarsen at late times in the following manner:
(i) $l_{+}(t)\approx {\sqrt {\pi t}}$
at late times and (ii) $l_{-}(t) \sim t^{1/2}{\log (bt)}$ where $b$ is a
number depending on the initial volume fraction of the `$+$' spins
which we will calculate explicitly (see below).
In fact, the main result we show below is that the ratio of the two length scales
behaves at late times as,
\begin{equation}
{l_{-}\over {l_{+}}} = {1\over {\epsilon}}{{\log (bt/t_0)}\over {\log b}}-1
\label{ratio}
\end{equation}
where $\epsilon$ is the initial volume fraction
(need not be small) and $t_0$ is some initial
time after which scaling starts holding. The equation (\ref{ratio}) explicitly
reflects the effect of broken symmetry.
Thus due to the dynamically generated local bias, the `$-$' domains grow
slightly faster than the `$+$' domains.
We also point out that in
contrast to the
spin models studied in the context of glassy systems\cite{Fred,SE},
the dynamics in our model does not freeze at zero temperature, rather
the domains coarsen indefinitely in an infinite system.

(2) Consequently, the magnetization $m(t)=(l_+ - l_-)/(l_+ + l_-)$ decays at late times
as,
\begin{equation}
m(t) = -1 + { {2\epsilon {\log b}}\over {\log (bt/t_0)} }.
\label{mag}
\end{equation}

(3) Evidently the number of `$+$' domains is same as the number of of `$-$' domains
since they alternate on the $1$-d lattice. The density
of domains of either `$+$' or `$-$' type per unit length: $N(t) = 1/[l_{-}+l_{+}]$ decays as,
\begin{equation}
{N(t)\over {N(t_0)}}= \sqrt{{t_0\over t}}\, {{\log b}\over {{\log (bt)}}}.
\label{domain}
\end{equation}

We also compute explicitly two other densities that play a somewhat central role
in our analysis:

(4) $r_1(t)$:  Given that a `$-$' domain has occurred, the probability
that it is of length $1$. We show below that
\begin{equation}
r_1(t) = {{\sqrt{\pi}}\over {\sqrt{t}} \log(bt/t_0)}
\label{r1}
\end{equation}
Note that the amplitude $\sqrt{\pi}$ is universal and independent of
volume fraction.

(5) $p_1(t)$: Given that a `$+$' domain has occurred, the probability that
it is of length $1$. We show that
\begin{equation}
p_1(t) = {1\over 2t} + {1\over t \log(bt/t_0)}
\label{p1}
\end{equation}
Once again the amplitude $1$ of the correction term is also universal
and independent of volume fraction.

The constant $b$ in the above equations can also be computed exactly,
and we show that,
\begin{equation}
b = \exp\left( {{\sqrt{\pi}}\over r_1(t_0)\sqrt{t_0}}\right)
\label{b}
\end{equation}

\section{Connection to Granular Compaction and Reaction Diffusion Systems}
We now establish a one to one mapping between our spin model (with $\alpha=0$)
and a simple lattice model of granular compaction.
Let us consider a $(1+1)$-dimensional granular packing where the grains
are represented by square blocks. The pack consists of horizontal
layers consisting of blocks and voids (see Fig 1).
We focus on the `active' layer, i.e., the first horizontal
layer that is not fully compact with blocks as we go up from the bottom
of the pile. Below this active layer, all layers are compact and remain compact
under vertical tapping, i.e., their dynamics is completely frozen. All
the activities take place in or above the `active' layer. We identify this
`active' layer with the one dimensional
lattice of the spin model. This active layer consists of sequences
of blocks (particles) and voids (holes). We identify a unit block or a particle
in the active layer as a `$-$' spin of our spin model. Similarly a hole
is identified as a `$+$' spin of the spin model.

As the system is tapped vertically, the particles in the active layer
can undergo the following moves: (1) In the interior of a row of consecutive
particles in the active layer, there is no effect of tapping as the system
is completely jammed there. The effect of vertical tapping is felt only at the edges
of a particle cluster. The particles at the edges, under tapping, can move up
or `tap' up
from the active layer to the layers above the active layer (see Fig. 1). However
if the cluster consists of an isolated particle sandwiched between two holes,
it has no place to move up, so it stays at its original location. (2) Tapping
also can `roll' off a particle residing in a layer above the active layer, into the
active layer, at the edges of the supporting cluster (see Fig. 1).

Let us now see what the rates of spin flips in Eq. (\ref{eq:rates}) in the spin model
imply for the compaction model. The rate $W(+;++) =0$ implies that if we have
three consecutive holes in the active layer, then a particle can not be
deposited in the middle site under tapping. This is because a new particle can appear
into the active layer only at the edges of the clusters of particles, but not
in the middle of a cluster of voids. The rate $W(-;--) = 0$ implies that
if we have three consecutive particles in the active layer, the middle particle
can not move up under tapping since it is completely jammed.
The rate $W(+;-+)=1/2$ signifies that if we have three consecutive sites in the active layer
consisting respectively of a particle, hole and a hole, then a particle
can appear in the middle hole with rate $1/2$. This is `rolling' off move to the right at
the edge of a cluster, as discussed in the previous paragraph. Similarly $W(+;+-)=1/2$
signifies the `rolling' off move to the left. The rate $W(-;+-)=1/2$ implies
that when we have three consecutive sites in the active layer consisting respectively
of a hole, particle and a particle, the middle particle can disappear (into the layers
above) with rate $1/2$. This is the `tapping' up move to the right from the edge of a cluster
as discussed in the previous paragraph. Similarly $W(-;-+)=1/2$ signifies
the rate of `tapping' up to the left from the edge of a cluster. The
rate $W(+;--) = 1$ indicates that if we have a sequence of particle, hole, particle in
the active layer, the middle hole can be filled up with a particle with rate $1$.
This is simply the addition of two `rolling' off rates from the left and the right
of the middle hole. Finally the rate $W(-;++) = \alpha =0$ implies that
if we have a sequence of hole, particle, hole in the active layer, the middle
particle can not disappear. This is because an isolated particle has no place to
move or `tap' up as discussed in the previous paragraph.
This last rate indeed breaks the particle hole symmetry.

Note that the rates $ W(+;-+) = W(+;+-) = W(-;+-)
= W(-;-+) = {1\over 2}$ correspond to the diffusion
of domain walls in the spin model. In the compaction
model, these are indeed the moves induced by the mechanical tapping.
If these rates were zero, i.e., no tapping, then the dynamics
in the active layer would freeze after all the isolated holes are
filled up and the system will be stuck in a metastable
configuration and hence the compaction will stop. It is these diffusion moves that lift
the system out of a metastable configuration and the system continues
to compactify, though extremely slowly. Identifying the `-' (+) spins
with particles (holes) in the granular model, it is easy to see
that the average magnetization $m$ in the spin model is related to
the average particle density $\rho(t)$ in the granular model via the
simple relation, $m(t)=1-2\rho(t)$. Thus the result in Eq. (\ref{mag}) for the magnetization
indicates that the density will grow to its fully compact value $1$ as
\begin{equation}
\rho(t) = 1-{ {\epsilon {\log b}}\over {\log (bt/t_0)} },  
\label{rhot}
\end{equation}
at very late times, as observed in
experiments on mechanically tapped granular media\cite{GE}. 

Other one dimensional lattice models, notably the `car parking' model,
has been previously used to
explain the logarithmic relaxation in granular materials\cite{CP}.
The local rules for the dynamics of particles in the car parking
model are, however, quite different from those in our model.
In the parking model, the rules for the particle motion are:
(i) a new particle can be  absorbed only at
sites containing
a hole at a rate proportional to the  number of neighboring particles and
(ii) a particle evaporates with a small rate (thus leaving behind a hole) if
it has one and only one neighboring hole. While this model also exhibits
an inverse logarithmic growth of the density, the dynamical moves
(in particular the fact that the desorption rate is infinitesimally small)
are chosen in a somewhat ad hoc manner. In contrast, as explained via 
the mapping detailed above, our model incorporates the basic 
minimal microscopic moves of the particles that are observed in the 
compaction  process.   

In terms of the motion of the domain walls between `$+$' and `$-$'
phases, our model can also be considered a new model of two
species reaction-diffusion in one dimension.
We note that in
the case $\alpha = 1$, the domain walls diffuse and annihilate upon
contact.  This corresponds to the process $A+A \to 0$ \cite{bray1}.
In the case $\alpha =0$, we need to distinguish between the two types
of domain walls $-+ \equiv A$ and $+- \equiv B$. Note that by
definition (originating from a spin configuration) the $A$'s and $B$'s
always occur alternately. Here both $A$'s and $B$'s diffuse as before;
however when an $A$ and a $B$ meet, they annihilate only if $A$ is
to the left of $B$, otherwise there is hard core repulsion between
them. We will show below that this hard core repulsion
between the particles is a relevant interaction
that changes the late time dynamics considerably. Recently the relevance
of hard core repulsion between particles in reaction-diffusion
systems have been explored in a number of contexts\cite{HC}.

\section{Some Exact Results}

To start with we write down two exact relations which are derived directly
from the Glauber dynamics (we shall see later that these exact relations
are in fact respected by the IIA approximation).

Let $N(t)$ be the number of domains of either `$+$' or `$-$' spins per unit
length. The density of kinks is therefore $2N(t)$.
Let $P_1$ be the density of the triplets `$-+-$' per unit length,
then clearly $p_1 = P_1/N$. We note also that the only way in which a
domain can be destroyed in an infinitesimal time step is by flipping
an isolated `$+$' spin in a triplet `$-+-$'. This gives
\begin{equation}
{dN\over dt} = -P_1.
\label{exactP:N}
\end{equation}
If $P(S_i)$ denotes the  probability 
that the spin at  site $i$ takes the
value $S_i$, the evolution of $P(S_i)$ depends only on the
rates in Eq. (\ref{eq:rates}) and the three point probability distribution
$P(S_{i-1},S_{i},S_{i+1})$.
The evolution of $P(S_i)$ is then given by
\begin{eqnarray}
{dP(S_i)\over dt} &=&
\sum_{S_{i+1}, S_{i-1}} W(-S_i; S_{i-1}
,S_{i+1}) P(S_{i-1}, -S_{i}, S_{i+1}) \nonumber
\\ &-& \sum_{S_{i+1}, S_{i-1}} W(S_i; S_{i-1}
,S_{i+1}) P(S_{i-1}, S_{i}, S_{i+1}) 
 \label{glauber}
\end{eqnarray}
Substituting $S_i = 1$ in the above equation we find
\begin{equation}
{dP(1)\over dt} = P(-1,-1,1) - P(-1,1,-1) - P(1,1,-1)
\end{equation}
where we have used the evident left to right symmetries $P(-1,-1,1)
= P(1,-1,-1)$ and $P(1,1,-1) = P(-1,1,1)$. We now observe that
$P(-1,-1,1) + P(1,-1,1)  = P(-1,1)$  and $P(-1,1,-1)+P(1,1,-1) = P(1,-1)
= P(-1,1)$ and thus find
\begin{equation}
{dP(1)\over dt} = - P(1,-1,1) .
\end{equation}
We note that $P(1,-1,1)$ is simply the probability that at a given site
the spin is a `$-$' spin and its two neighbors are `$+$' spins, that is to
say that their is a `$+-+$' defect at the site considered.
If we now sum this equation over each site on an interval of unit length
on the lattice and recall that $R_1$ is the density of the `$+-+$' triplets per
unit length, we obtain
\begin{equation}
{dL_+\over dt} = - R_1.
\label{exactR:L}
\end{equation}
where $L_+$ is the fraction of the `$+$' spins.
Note that  $L_+ + L_- = 1$, where $L_-$ is the fraction of the `$-$'
spins. From the relation Eq. (\ref{exactR:L}) we obtain Eq. (\ref{eq:em})
for the evolution of the average magnetization $m$. Physically it is easy to
see the origin of Eq. (\ref{exactR:L}) as,  on an average,
the fraction of `$+$' spins can decrease only due to the blockage
by `$+-+$' triplets.

Writing Eq. (\ref{exactR:L}) in terms of $r_1 = R_1/N$ and
the average length of the `$+$' domains $l_+ = L_+/N$ we obtain
\begin{equation}
{\dot N \over N}  = -{\dot l_+ + r_1\over l_+},
\label{dN/dt}
\end{equation}
where $\dot{x} = dx/dt$.
This equation can be integrated starting at some arbitrary time $t_0$ to
give
\begin{equation}
{N(t) \over N(t_0)} = {l_+(t_0)\over l_+(t)}\exp\left(-\int_{t_0}^t
{r_1(t')\over l_+(t')}dt' \right).
\label{Nsol1}
\end{equation}
Furthermore, if the volume fraction of the `$+$' phase is, $L_{+}(t_0)=\epsilon$, then, using the relation,
$N(t) = 1/[l_{-}(t) + l_{+}(t)]$ in Eq. (\ref{Nsol1}), we find,
\begin{equation}
{l_-(t)\over l_+(t)} = {1\over \epsilon}\exp\left(\int_{t_0}^t
{r_1(t')\over l_+(t')}dt' \right) - 1,
\label{ratio1}
\end{equation}
clearly showing that the ratio $l_-(t)/l_+(t)$ is growing due to the presence of the
triplets `$+-+$'. Note that the asymmetry between the growth of `$-$' and `$+$' domains
is evident due to the presence of the triplet defects `$+-+$' with density $R_1=r_1 N$.

All the results presented above are exact. To derive the late time behavior of the model
we first consider below the IIA. We solve the IIA equations self-consistently and show
that the IIA precisely predicts the results mentioned in Section II.
Besides we shall argue that in the case where the initial
volume fraction of the `$+$' domains is small, i.e.,
$\epsilon \ll 1$,
correlations do not develop between the domains if no correlations are
present in the initial conditions and hence the IIA is exact to
leading order in $\epsilon$. At the end we present a very simple heuristic argument which is also in agreement with these results.

\section{IIA analysis}

In this section we consider the  IIA where correlations
between neighboring domains are neglected. The IIA was used previously
for the $\alpha=1$ case\cite{KB} yielding results in agreement,
qualitatively as well as quantitatively to a fair degree of accuracy,
with the exact results available \cite{glauber,DZ}.  Let $P_n(t)$ and
$R_n(t)$ denote respectively the number density of `$+$' and `$-$'
domains of length $n$ at time $t$. Note that $R_1(t)$ is the density of the
triplet `$+-+$' as before.
Let $N(t)=\sum_n P_n=\sum_n R_n$
denote the domain density of `$+$' or `$-$' spins. Also the fractions
$L_{\pm}(t)$ of `$+$' and `$-$' spins are given by, $L_{+}(t)=\sum_n
nP_n$ and $L_{-}(t)=\sum_n nR_n$ with $L_+(t) + L_-(t) = 1$.
During an infinitesimal time step
$\Delta t$, $P_n(t)$ evolves as:
\begin{eqnarray}
P_n(t + \Delta t) &= P_n(t) - \Delta t P_n(t) - \Delta t P_n(t)
\left[1 - {R_1(t) \over N(t)} \right] \nonumber \\ &+ \Delta t
P_{n+1}(t) + \Delta t P_{n-1}(t)\left[1 - {R_1(t) \over N(t)} \right].
\label{deltaPn}
\end{eqnarray}
The right hand side of the above equation includes the
various loss and gain terms. The second and the third terms describe
respectively the loss due to the hopping inward and hopping outward of
the domain walls at the two ends of a `$+$' domain of size $n$. An
outward hop can occur provided the neighboring domain in the direction
of the hop is not an isolated `$-$' spin and this is
ensured by the prefactor $(1 - R_1/N)$ in the third term. The fourth
and the last term describe similarly the corresponding gains.
One can similarly write down the evolution equation for the
$R_n(t)$'s. During an infinitesimal time step
$\Delta t$, $R_n(t)$ evolves as:
\begin{eqnarray}
R_n(t + \Delta t) &=& R_n(t) - \Delta t R_n(t) - \Delta t R_n(t)
\left[1 - {P_1(t) \over N(t)} \right]  \nonumber \\
&-&2 \Delta t P_1(t){R_n(t) \over N(t)} + \Delta t R_{n+1}(t) \nonumber \\
&+& \Delta t R_{n-1}(t)
\left[1 - {P_1(t) \over N(t)} \right] \nonumber \\
&+& {P_1(t) \over N^2(t)}\sum_{i=1}^{n-2} R_i(t)R_{n-i-1}(t) \,\,\, ;
n \geq 2
\label{deltaRn}
\end{eqnarray}
and
\begin{eqnarray}
R_1(t + \Delta t) &=  R_1(t) - \Delta t R_1(t)\left[1 - {P_1(t) \over N(t)} \right] \nonumber \\
&- 2 \Delta t P_1(t){R_1(t) \over N(t)} + \Delta t R_2(t).
\label{deltaR1}
\end{eqnarray}
The negative (loss) terms in Eq. (\ref{deltaRn}) for domains of size 
$n\geq 2$ may be understood as follows.
A domain of length $n$ may be lost by the domain wall at either end jumping
inwards with rate $1/2$. This term is the second term 
in Eq. (\ref{deltaRn}) and  
as there are two  domain walls we have a factor of $2$. 
A domain of length $n$ may also be lost by a domain wall hopping
outwards.  This happens with rate $1/2$ 
if the neighboring domain is
not a triplet `$-+-$'.  The third term of Eq. (\ref{deltaRn})
corresponds to this event, the factor $(1 - P_1(t)/N(t))$ is the probabilty
of the absence of a  triplet `$-+-$' as a neighboring domain. There is
again a factor of $2$ coming from the fact that there are two domain walls. 
However if a neighboring domain  is of type `$-+-$' the outward jump 
towards this domain occurs with rate $1$ 
(as the central $+$ spin flips with rate $1$). The term corresponding to
these two events (from the right and left domain walls) is the fourth term
in  Eq. (\ref{deltaRn}). 
The two first gain terms come from
identical arguments and the last convolution term represents domain
coalescence, where a domain of length $n$ is formed with rate $1$ from
two `$-$' domains of length $i$ and $n-i-1$ (where $1\leq i \leq n-2$)
if they are separated by a `$-+-$' triplet. Eq. (\ref{deltaR1}) is
obtained in a similar fashion with the exception that the hard core
repulsion generates a reflecting boundary condition.
Taking the limit $\Delta t \to 0$ in the above equations we obtain
the IIA equations for the evolution of the domain densities
\begin{equation}
{dP_n\over dt} = P_{n+1} + P_{n-1} - 2 P_n + {R_1\over N}(P_n - P_{n-1})
\label{eq:P_n}
\end{equation}
for all $n\geq 1$ with $P_0=0$ (absorbing boundary condition) and
\begin{eqnarray}
{dR_n\over dt} &=& R_{n+1} + R_{n-1} - 2 R_n -{P_1\over N}(R_n+R_{n-1})  \nonumber \\
&\ & \ \ \ \ + {P_1\over N^2} \sum_{i=1}^{n-2}R_i R_{n-i-1} ; \ \ \ n\geq 2
\nonumber \\ {dR_1\over dt}&=& R_2-R_1 -{P_1\over N}R_1 ,
\label{eq:R_n}
\end{eqnarray}
It is
somewhat convenient to use the normalized variables $p_n = {P_n/ N}$
and $r_n = {R_n/ N}$. The average domain lengths are then
given by $l_+(t) = \sum np_n$ and $l_-(t) = \sum n r_n$ and the domain density
$N(t) = 1/(l_+(t) + l_-(t))$.
In terms of these normalized variables the IIA equations are given by,
\begin{equation}
{dp_n\over dt} = p_{n+1} + p_{n-1} - 2 p_n + r_1(p_n - p_{n-1}) + p_1
p_n
\label{eq:p_n}
\end{equation}
for all $n\geq 1$ with $p_0=0$ (absorbing boundary condition) and
\begin{eqnarray}
{dr_n\over dt} &=& r_{n+1} + r_{n-1} - 2 r_n -p_1 r_{n-1} \nonumber \\
&\ & \ \ \ \ + p_1 \sum_{i=1}^{n-2}r_i r_{n-i-1} ; \ \ \ n\geq 2
\nonumber \\ {dr_1\over dt}&=& r_2-r_1 .
\label{eq:r_n}
\end{eqnarray}
It is easy to check that the normalization condition, $\sum p_n=\sum r_n =1$ is satisfied by these two equations.
We note that Eq. (\ref{eq:p_n}) or more clearly its unnormalized version in 
Eq. (\ref{eq:P_n}), i.e., $\dot{P}_n = P_{n+1} + P_{n-1} - 2P_n +
r_1(P_n - P_{n-1})$, just represents the motion of a random walker on the
positive side of a $1$-d lattice with a sink at the origin ($P_0 = 0$)
and a time dependent drift term (proportional to $r_1$).
To calculate $N(t)$ using Eq. (\ref{Nsol1}),
we need to evaluate two quantities from the IIA equations: (i) $r_1(t)=R_1/N$ and (ii) $l_+(t)=\sum np_n$.

The two IIA equations above are coupled nonlinear equations with infinite number of variables and
hence exact solution of Eqs. (\ref{eq:p_n}) and (\ref{eq:r_n}) are difficult.
Our approach will be a combination of a
scaling assumption and then  rechecking this assumption for
self-consistency. Consider first the $r_n$ equation, i.e. Eq.
(\ref{eq:r_n}).
On the right hand side, we will first ignore the diffusion term, solve for the rest and show that indeed neglecting the
diffusion term was justified in the first place. This is self-consistency.
Ignoring the diffusion term, we have the following equation,
\begin{equation}
{dr_n\over dt} = p_1\left[\sum_{i=1}^{n-2}r_i r_{n-i-1}-r_{n-1}\right],
\label{ar_n}
\end{equation}
with the reflecting boundary condition, $r_2=r_1$.
It is now easy to see that Eq. (\ref{ar_n}) admits a scaling solution,
$r_n(t) \approx \lambda(t)\exp [-n\lambda(t)]$, where ${\dot {\lambda}}(t)=-p_1(t)\lambda (t)$. Using the exact relation, $dN/dt=-p_1N$,
we get, $\lambda(t)=\lambda(t_0)N(t)/N(t_0)$. Note that we still do not know what $N(t)$ is. Now let us substitute this
solution to estimate the diffusion term that had been neglected in the first place. Clearly, the diffusion term, $T_{diff}=r_{n+1}
+r_{n-1}-2r_n \sim O({\lambda^3(t)})$, whereas the other terms (for example the left hand side of Eq. (\ref{eq:r_n})) typically scale
as $\sim O({\dot \lambda}(t))\sim O(p_1(t)\lambda (t))$. Thus, in order to be self-consistent in neglecting the diffusion term, we need to have
$p_1(t)\gg {\lambda}^2(t)\sim N^2(t)$.
We will see that this condition is actually satisfied once we derive the expression for $N(t)$. This just means that the diffusion terms only
contribute to the corrections to the leading scaling behavior.

From the above analysis, we find to leading order for large $t$, $r_1\approx \lambda(t)=\lambda(t_0)N(t)/N(t_0)$, i.e.,
$r_1(t)\approx r_1(t_0)N(t)/N(t_0)$.
We now need to evaluate the other remaining quantity, $l_+(t)=\sum np_n$. For this we now turn to the $p_n$ equation,
Eq. (\ref{eq:p_n}). In this equation,
we will again first ignore the drift term $r_1(p_n-p_{n-1})$ solve for the rest and check that indeed the neglect of the drift term
was justified. Ignoring the drift term, we get,
\begin{equation}
{dp_n\over dt} = p_{n+1} + p_{n-1} - 2 p_n + p_1p_n.
\label{ap_n}
\end{equation}
with the absorbing boundary condition, $p_0=0$. This equation can be solved
exactly.
Indeed it also admits a scaling solution, $p_n(t)= t^{-1/2}f(nt^{-1/2})$,
where the scaling function (normalized to unity) is given by, $f(x)={x\over 2}\exp(-x^2/4)$.
Now let us estimate the drift term that was neglected.
Clearly the drift term, $r_1(p_n-p_{n-1})\sim O(r_1/t)\sim O(N(t)/t)$ since $r_1\sim N(t)$ from previous paragraph.
The other terms in the Eq. (\ref{eq:p_n})
(for example the left hand side of Eq. (\ref{eq:p_n})) is of order, $t^{-3/2}$ at late times.
Thus, for self-consistency in neglecting the drift term, we need to show that
$t^{-3/2}\gg N(t)/t$. We will again see that this condition
is indeed also satisfied once we obtain the
expression for $N(t)$. From this form of $p_n(t)$, we thus obtain,
to leading order for large $t$, $l_+(t)=\sum np_n \approx t^{1/2}\int_0^{\infty}xf(x)dx$.
Using $f(x)={x\over 2}\exp(-x^2/4)$, and doing the integral we finally find,
$l_+(t)\approx \sqrt {{\pi t}}$ for large $t$.

Using these two results (i) $r_1(t)=r_1(t_0)N(t)/N(t_0)$ and (ii) $l_+(t)=\sqrt{\pi t}$ in the
exact equation Eq. (\ref{Nsol1}) and
differentiating with respect to $t$,
we find a differential equation for $N(t)$,
\begin{equation}
{d\over dt}\left( \sqrt{t} N(t) \right) =
- {r_1(t_0)\over\sqrt{\pi} N(t_0)}N^2(t).
\label{Nsol2}
\end{equation}
Introducing the dimensionless variable
\begin{equation}
S(t) = \sqrt{{t_0\over t}}\,{N(t)\over N(t_0)}
\label{eqS}
\end{equation}
in Eq. (\ref{Nsol2}) we find
\begin{equation}
{dS\over dt} = -\log(b) {S^2\over t}
\label{Nsol3}
\end{equation}
where $\log(b) = {\sqrt{\pi}\over r_1(t_0)\sqrt{t_0}} $.
Integrating Eq. (\ref{Nsol3}) we find
\begin{equation}
S(t) = {\log(b)\over \log(bt/t_0)},
\label{Ssol}
\end{equation}
we thus obtain the result
\begin{equation}
{N(t)\over N(t_0)} = \sqrt{{t_0 \over t}} \,{\log(b)\over \log(bt/t_0)}.
\label{Nfinal}
\end{equation}
Substituting this result in the expression, $r_1(t)\approx r_1(t_0)N(t)/N(t_0)$, we get
\begin{equation}
r_1(t) = {\sqrt{\pi}\over \sqrt{t} \log(bt/t_0)}.
\label{r1final}
\end{equation}
Next we use the late time result Eq. (\ref{Nfinal}) in the exact relation Eq.
(\ref{exactP:N}) and find,
\begin{equation}
p_1 = {1\over 2 t} + {1\over t \log(bt/t_0)}.
\label{p1final}
\end{equation}

Let us check the two self-consistency conditions, (a) $p_1(t)\gg {\lambda}^2(t)\sim N^2(t)$ and (b) $t^{-3/2}\gg N(t)/t$.
Using the expression for $p_1$ from Eq.
(\ref{p1final}) and that of $N(t)$ from Eq. (\ref{Nfinal}), it is immediately evident that indeed these two conditions
are satisfied for large $t$.
Thus our whole approach has been completely self-consistent and the IIA results are precisely those mentioned in the abstract.
Note also that $t_0$ must be sufficiently large such that both scaling laws, $r_1(t)\sim N(t)$ and
$l_+(t)\sim {\sqrt {\pi t}}$ start holding for $t >t_0$.

\section{Zero Volume Fraction Limit}

In this section, we show that the IIA results essentially become exact in the zero volume fraction
limit of the `$+$' phase, i.e., in the limit $\epsilon \to 0$. Suppose we start from an initial condition
such that, $r_n(0)=\epsilon (1-\epsilon)^n$ and $p_n(0)=\delta_{n,1}$. This means that in the initial condition,
the average length of the `$-$' domains, $l_-(0)\sim 1/{\epsilon}$, whereas $l_+(0)=1$. Thus the `$-$' domains
are typically much larger than the `$+$' domains, in the limit $\epsilon\to 0$. Also initially all the domains are
completely uncorrelated. So the picture is as follows. We have little droplets of `$+$' phase in a sea of `$-$' phase. Besides,
one can also compute the initial density of domains of either `$+$' or `$-$' types. It is given by, $N(0)=1/{\epsilon (1-\epsilon)}
\sim 1/{\epsilon}$ to leading order in $\epsilon$.

Now let us consider the time evolution of the system starting from this initial condition. As time increases, the
`$+$' domains will
certainly grow in size. But a typical `$+$' domain will disappear (via the absorbing boundary condition) much before encountering
other `$+$' domains, i.e., before feeling the presence of the constraint due to triplets `$+-+$'. The probability of such an event
is of order $O(\epsilon)$. Thus effectively, the dynamics of the system will proceed via eating up of the `$+$' domains. Hence,
if there is no correlation between domains in the initial condition, the dynamics is not going to generate correlations between them.
This is precisely what happens in the zero temperature dynamics of the $q$ state Potts model in $1$-d in the limit $q\to 1^{+}$\cite{CM,KB}.

Thus in this limit, the evolution of the `$-$' domains is governed by the exact equation,
\begin{equation}
{dr_n\over dt} = p_1\left[\sum_{i=1}^{n-2}r_i r_{n-i-1}-r_{n-1}\right],
\label{epsr_n}
\end{equation}
which is same as the IIA equation Eq. (\ref{eq:r_n}) without the diffusion term. Starting from the initial condition,
$r_n(0)=\epsilon(1-\epsilon)^n$, one can solve the above equation for any $t$ exactly to leading order in $\epsilon$.
It turns out that, to leading order in $\epsilon$, Eq. (\ref{epsr_n}) admits a solution, $r_n(t)= \mu (t)[1-\mu
(t)]^{n-1}$,
where $\mu(t)=\epsilon \exp [-\int_0^t p_1(t')dt']$. Using once again, the exact equation $dN/dt=-p_1N$, we find,
$\mu (t) =\epsilon N(t)/N(0) +O(\epsilon^2)= N(t) +O(\epsilon^2)$ where we have used $N(0)=1/{\epsilon} +O(\epsilon^2)$. Thus we get,
\begin{equation}
r_1(t)=\mu (t)= N(t) + O(\epsilon^2).
\label{epsr_1}
\end{equation}

Now, let us consider the evolution of the `$+$' domains. Since a typical `$+$' domain never encounters (to leading order in $\epsilon$)
any other `$+$' domain and hence does not feel the constraint due to $r_1$'s, the effective dynamics of a `$+$' domain is
that of a single `$+$' domain immersed in a sea of `$-$' phase. Let $P_n$ denote the probability that such a domain is of length $n$.
Then, to leading order in $\epsilon$,  $P_n$'s clearly evolve by the simple diffusion equation,
\begin{equation}
{dP_n\over dt} = P_{n+1} + P_{n-1} - 2 P_n,
\label{diff}
\end{equation}
with the absorbing boundary condition, $P_0=0$. The normalized conditional probability, $p_n=P_n/N$ with $N=\sum P_n$, then satisfies the
equation,
\begin{equation}
{dp_n\over dt} = p_{n+1} + p_{n-1} - 2 p_n + p_1p_n ,
\label{epsp_n}
\end{equation}
same as Eq. (\ref{ap_n}). This equation has to be solved with the initial condition, $p_n(0)=1$. It can be solved exactly. Without writing the
explicit solution, we just mention the result for $\sum np_n$. We find that for large $t$ and leading order in $\epsilon$,
\begin{equation}
l_+(t)=\sum np_n\approx \sqrt {\pi t} + O(\epsilon).
\label{epsl+}
\end{equation}

Using the results from Eqs. (\ref{epsr_1}) and (\ref{epsl+}), i.e., (i) $r_1(t)=N(t)+ O(\epsilon^2)$
and (ii) $l_+(t)\approx \sqrt {\pi t} +O(\epsilon)$ in the exact equation Eq. (\ref{Nsol1}), we once again recover all
the IIA results of the previous section , with $b={\sqrt \pi}/{\epsilon}$.

Hence IIA becomes exact in the $\epsilon\to 0$ limit. This is not surprising as the dynamics in this limit
does not generate correlations if there are none in the initial condition.

\section{ Other Volume Fractions}

For finite initial volume fraction of the `$+$' phase, the IIA can not be exact since the diffusion of kinks
correlate the domains as time progresses, even if the domains had no correlations to start with.
However, the volume fraction of
the `$+$' phase decreases monotonically with time according to the exact equation Eq. (\ref{exactR:L}). Thus at very late times
when the volume fraction $L_+(t)$ is very small, the effective fixed point picture of the system is very similar to the $\epsilon\to 0$
limit picture, i.e., small `$+$' domains immersed in the sea of `$-$' domains. The only difference is that the big `$-$' domains
may now be correlated. However, it is very likely (though we can not prove this rigorously) that the correlations between
domains
are very small at very late times and therefore the IIA results ((1)-(4) in the abstract) become asymptotically exact.
The numerical results reported in \cite{mdg} appear to confirm this
fact.

Actually if this late time fixed point picture is correct, then one can derive all the results from a very simple
heuristic argument, presented below.

\section{Heuristic Approach}

The heuristic picture is as follows. Due to the facts that
`$+$' domains can grow only by diffusion and the `$-$' domains grow by
diffusion and coalescence, one expects that at late times the ratio
$l_+(t)/l_-(t) \to 0$; this is also clear from Eq. (\ref{exactR:L}).
Therefore at late times one expects to find  `$+$' domains sandwiched
between much larger `$-$' domains. Consequently  the `$+$'
domains, to a first approximation, never encounter the `$+-+$' triplets
at late times and
hence diffuse freely and annihilate via the disappearance of `$-+-$' triplets.
Therefore the length of a  $+$' domain behaves as a diffusion with a killing
boundary condition at the origin. Solving the discrete diffusion equation
corresponding to this picture (see Eq. (\ref{epsp_n})), one finds that at
late times
\begin{equation}
l_+(t) \approx \sqrt{\pi t}
\label{diff+}
\end{equation}
If one now views the system at the length scale of the
`$-$' domains $l_-(t)\gg l_+(t)$, one sees long stretches of `$-$' domains occasionally interrupted by `$+$' domains
which are now shrunk to a single point when viewed from the length scale of the `$-$' domains.
The rate of occurrence of these points per unit length $\lambda(t)$ is clearly proportional
to $N(t)$, i.e., the kink density. Note that $r_n$ is simply the conditional probability: Given that
a `$-$' domain has occurred, what is the probability that it is of length $n$. Now
if one assumes
that these punctual `$+$' domains (or the points) are distributed randomly, one finds that $r_n$ is simply
given by the geometric distribution,
$r_n = \lambda(t)(1-\lambda(t))^{n-1}$ where $\lambda(t) = c N(t)$ for some
constant $c$. We therefore find
\begin{equation}
r_1(t) \approx c N(t),
\label{eq:r1approx1}
\end{equation}
for late times. Furthermore, if we denote by $t_0$ a large time after which
this picture becomes valid we may write
\begin{equation}
r_1(t) \approx r_1(t_0) {N(t)\over N(t_0)}.
\label{r1approx2}
\end{equation}

These two expressions for $r_1$ and $l_+(t)$ once again are same as obtained
by a more careful analysis of the IIA equations
and when substituted in Eq. (\ref{Nsol1}), they give the same IIA results once again. Thus the basic assumption of this
heuristic picture is the `$+$' domains occur randomly, which seems like an 
accurate description at late times.

\section{Conclusions}

In this paper we have presented detailed analytical studies on a simple
one dimensional kinetically constrained Ising model which was introduced
in Ref.\cite{mdg}. The kinetic constraints in this model are local
and dynamically generated. The effect of these constraints was shown
to slow down the dynamics rather drammatically. We have shown that
the average magnetization in this model decays extremely slowly
with time in an inverse logarithmic fashion to its final saturation value.
This kind of inverse logarithmic law was observed in 
the behavior of the density of granular material in experiments
on granular compaction\cite{GE} and was also seen in numerical
simulations of various lattice based and `tetris' like models\cite{NI}. There
have been some theoretical arguments proposing various mechanisms
responsible for this slow compaction\cite{FV,CP}. These include
the free volume argument\cite{FV} and the argument based on an analogy
with car parking models with the  somewhat ad hoc assumption that the
cars `depark' from a lane at an infinitesimal rate\cite{CP}. In contrast, in this paper
we have mapped our kinetic Ising model {\em directly} to a lattice model
of granular compaction which incorporates the basic minimal microscopic moves
in the compaction process. The average magnetization $m(t)$ in the Ising model,
via this mapping, gets related to the density of compaction $\rho(t)$
in the granular model as $\rho(t)=(1-m)/2$.   
Hence, besides having nontrivial behavior
and yet analytically solvable, our toy model of granular compaction
correctly reproduces the inverse logarithmic time dependence seen 
in the experiments\cite{GE} and thereby proposes a new 
and entirely different mechanism 
for this slow compaction, quite different from the previous models
such as the car parking models. It is also interesting to note a 
study of compaction in the Tetris model \cite{BKLR} shows that at late times
the activity of the system, leading to compaction, occurs at boundaries 
between domains which can be identified in the system. The image of 
compaction as a kinetically hindered coarsening process thus appears to be
quite robust.    

From a somewhat broader perspective, our work addresses a general question:
what is the effect of kinetically generated disorders on the coarsening
dynamics in domain growth problems? In the present work
we have studied a specific type of kinetic disorder, namely a
dynamically generated local magnetic field. This field
acts locally on the topological defects responsible for the
coarsening process (in this case simple domain walls). Our study
suggests that such kinetic disorders, while slowing down the dynamics drastically,
do not altogether inhibit the coarsening process
as found in other constrained kinetic Ising models\cite{Fred}. 
The domain growth problems are rather common and occur in various 
physical systems\cite{bray}. 
Our work, therefore,  opens up the possibility 
of studying the slowing down
in coarsening dynamics due to kinetic disorders in many of these
systems. 
For example, it would be interesting to study the effect of dynamically generated local fields 
in higher dimensionional Ising models,
in $O(n)$ vector spin models and  in liquid crystals, to mention a few. 
It is possible to have other types of local kinetic disorders than the one studied here
and it would also be interesting to study their effect  
in coarsening systems.

We thank P. Grassberger for his earlier collaboration in this 
study, and also M. Barma for useful discussions. We acknowledge
interesting exchanges with G. Odor, H. Hinrichsen
and G. Sch\"utz.

\end{multicols}

{\bf {Figure Captions}}

Fig. 1. Picture of rolling off (left domain) and tapping up (right domain)
        in the granular interpretation of the spin model, the solid
        squares represent particles.

\begin{figure}
\narrowtext
\epsfxsize=1\hsize
\epsfbox{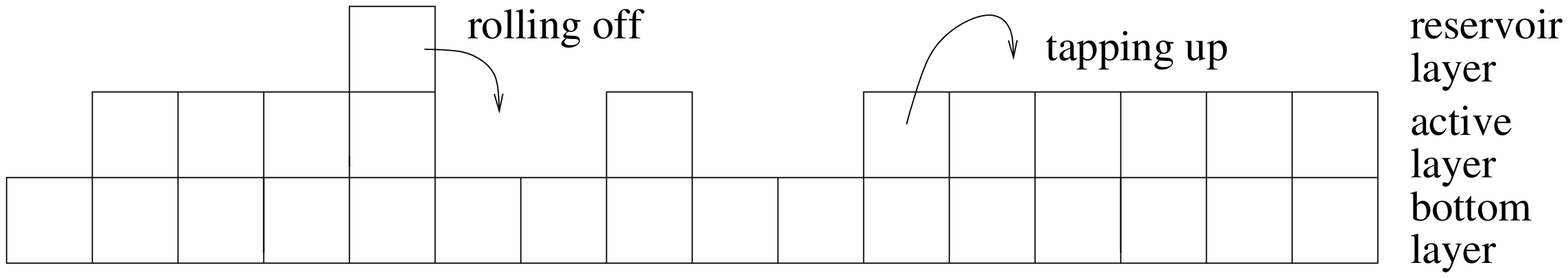}
\label{rollingfig}
\end{figure}

\end{document}